# О затухании видеоимпульса в нелинейных линиях передачи с идеальным диэлектриком

*А. С. Кюрегян*

Всероссийский Электротехнический институт им. В. И. Ленина, 111250, Москва, Россия
E-mail: semlab@yandex.ru

Затухание видеоимпульса с монотонно нарастающим напряжением $U_0(t)$ на входе в линию передачи с идеальным диэлектриком можно охарактеризовать «омическим» падением напряжения $U_\sigma(t)$ вдоль электродов с конечной проводимостью $\sigma$. Для вычисления $U_\sigma(t)$ в коаксиальной и полосковой линиях получены точные аналитические формулы без учета и с учетом сильного скин-эффекта, которые не зависят от дисперсии и степени нелинейности диэлектрика и поэтому пригодны для оценки затухания ударных электромагнитных волн.

Пусть на вход незаряженной коаксиальной линии передачи с идеальным диэлектриком[1] подается монотонно нарастающий видеоимпульс напряжения $U_0(t)$. Распространение сигнала при известных [1] условиях описывается системой телеграфных уравнений

$$\frac{\partial U}{\partial z} + L\frac{\partial I}{\partial t} + E_\sigma = 0, \qquad (1)$$

$$\frac{\partial I}{\partial z} + \frac{\partial Q}{\partial t} = 0, \qquad (2)$$

с начальным и граничным условиями

$$U(0,z) = 0 = I(0,z), \quad U(t,0) = U_0(t), \qquad (3)$$

где $U = U(t,z)$ и $Q = Q(t,z)$ - разность потенциалов между электродами и линейная плотность заряда, $E_\sigma = E_\sigma^+(t,z) - E_\sigma^-(t,z)$, $E_\sigma^\pm(t,z)$ - тангенциальные компоненты напряженности поля на границах между диэлектриком и электродами, обусловленные протеканием вдоль линии тока $I = I(t,z)$, $L = \frac{\mu_0}{2\pi}\ln\frac{r_e}{r_i}$ - погонная индуктивность, $\mu_0$ магнитная постоянная, $r_{e,i}$ - внешний и внутренний радиусы диэлектрической трубки.

Будем считать, что толщина электродов много больше $2\sqrt{D_\sigma t} \sim \sqrt{t/1\,\text{ns}} \cdot 8\mu m$, а ток $I(t,z)$ мало изменяется на расстоянии порядка $2\sqrt{D_\sigma t}$, где $D_\sigma = 1/\mu_0\sigma \sim 160\,\text{cm}^2/\text{s}$ – коэффициент магнитной диффузии, $\sigma$ - проводимость электродов. В этом случае оправдано приближение сильного планарного скин-эффекта и, используя теорию диффузии магнитного поля в металле (см., например, [2]), нетрудно получить соотношение

$$I(t,z) = \frac{\sqrt{t_\sigma}}{\pi L}\int_{t_z}^{t} E_\sigma(\vartheta,z)\frac{d\theta}{\sqrt{t-\theta}}, \qquad (4)$$

где $t_\sigma = \frac{\pi}{D_M}\left(\frac{2r_e r_i}{r_e+r_i}\ln\frac{r_e}{r_i}\right)^2$, $t_z = t_f(z)$ - время достижения передним фронтом сигнала плоскости $z$. Интегрирование (4) по $z$ приводит к уравнению Абеля [3]

---

[1] Мы называем идеальным диэлектрик, проводимость которого равна нулю.

$$\int_0^t U_\sigma(\theta)\frac{d\theta}{\sqrt{t-\theta}} = \frac{\pi L}{\sqrt{t_\sigma}}\int_0^{z_t} I(t,z)dz \equiv \Phi(t) \tag{5}$$

для «омического» падения напряжения $U_\sigma(t) = \int_0^{z_t} E_\sigma(t,z)dz$ на обоих электродах между входом линии и передним фронтом сигнала $z_t = z_f(t)$, дальше которого $U = 0 = I$. При монотонно нарастающей функции $U_0(t)$ напряжение $U_\sigma(t)$ характеризует затухание видеоимпульса в линии.

Так как $\Phi(0) = 0$, то решение (5) имеет вид [3]

$$U_\sigma(t) = \frac{1}{\pi}\int_0^t \frac{d\Phi(\theta)}{d\theta}\frac{d\theta}{\sqrt{t-\theta}}. \tag{6}$$

Из определения $\Phi(t)$ и уравнения (1) следует, что

$$\frac{d\Phi(t)}{dt} = \frac{\pi}{\sqrt{t_\sigma}}\left[I(t,z_t)L\frac{dz_t}{dt} - U(t,z_t) + U_0(t) - U_\sigma(t)\right], \tag{7}$$

где $dz_t/dt$ - скорость распространения фронта видеоимпульса. Во всех реалистичных случаях два первых слагаемых в правой части (7) равны нулю, поэтому равенство (6) является уравнением Абеля второго рода

$$\int_0^t U_\sigma(\theta)\frac{d\theta}{\sqrt{t-\theta}} + \sqrt{t_\sigma}\,U_\sigma(t) = \int_0^t U_0(\theta)\frac{d\theta}{\sqrt{t-\theta}} \tag{8}$$

для функции $U_\sigma(t)$. Его решение имеет вид [3]

$$U_\sigma(t) = F(t) + \frac{\pi}{t_\sigma}\int_0^t F(\theta)\exp\left(\pi\frac{t-\theta}{t_\sigma}\right)d\theta, \tag{9}$$

$$F(t) = \frac{1}{\sqrt{t_\sigma}}\int_0^t U_0(\theta)\frac{d\theta}{\sqrt{t-\theta}} - \frac{1}{t_\sigma}\int_0^t\left[\int_0^\theta U_0(\tau)\frac{d\tau}{\sqrt{\theta-\tau}}\right]\frac{d\theta}{\sqrt{t-\theta}} \tag{10}$$

В качестве примера рассмотрим случай трапецеидального нарастания напряжения на входе в линию

$$U_0(t) = V\begin{cases} t/t_0 & \text{при } t < t_0, \\ 1 & \text{при } t > t_0, \end{cases} \tag{11}$$

для которого получается

$$F(t) = \frac{V}{t_0}\left\{\frac{4}{3\sqrt{t_\sigma}}\left[t^{3/2} - (t-t_0)^{3/2}H(t-t_0)\right] - \frac{\pi}{2t_\sigma}\left[t^2 - (t-t_0)^2 H(t-t_0)\right]\right\}, \tag{12}$$

где $H(t)$ - ступенчатая функция Хевисайда. Зависимости $F(t)$ и $U_\sigma(t)$ для этого случая изображены на Рис. 1. Видно, что второе слагаемое в правой части (9) пренебрежимо мало в актуальном диапазоне значений $U_\sigma < 0.3V$, поэтому для расчета $U_\sigma(t)$ можно использовать простую формулу $U_\sigma(t) \approx F(t)$. В простейшем случае ступенчатого сигнала или при $t_0/t \to 0$

$$F(t) = V\left(2\sqrt{\frac{t}{t_\sigma}} - \pi\frac{t}{t_\sigma}\right) \tag{13}$$

и для $U_\sigma(t)$ получается формула

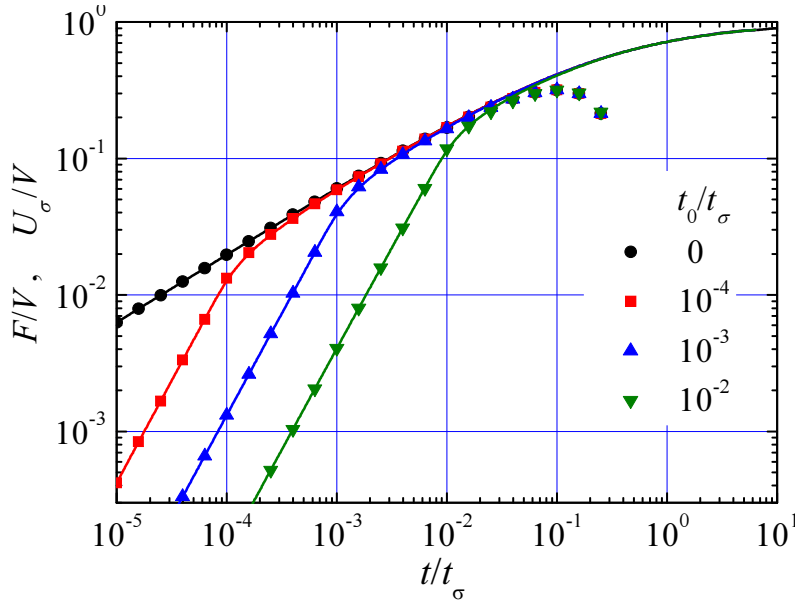

Рис. 1. Зависимости $U_\sigma(t)$ (линии) и $F(t)$ (символы), рассчитанные по формулам (9), (11) и (12) для четырех значений отношения $t_0/t_\sigma$

$$U_\sigma(t) = V\left[1 - \exp\left(\pi \frac{t}{t_\sigma}\right)\text{erfc}\sqrt{\pi \frac{t}{t_\sigma}}\right]_{t \ll t_\sigma} \approx 2V\sqrt{\frac{t}{t_\sigma}}, \qquad (14)$$

где $\text{erfc}(x)$ - дополнительная функция ошибок. Формулы (9)-(14) пригодны и для полосковой линии передачи, ширина которой много больше расстояния $d$ между электродами, однако в этом случае следует положить $t_\sigma = \pi d^2/D_M$.

Если толщина электродов много меньше $2\sqrt{D_\sigma t}$, то можно пренебречь скин-эффектом и переписать уравнение (1) в виде

$$\frac{\partial U}{\partial z} + t_R \frac{\partial E_\sigma}{\partial t} + E_\sigma = 0, \qquad (15)$$

где $t_R = L/R$ и $R$ - сумма погонных сопротивлений электродов. Легко показать, что в этом случае

$$U_\sigma(t) = \int_0^t U_\sigma(\theta)\exp\left(\frac{\theta - t}{t_R}\right)\frac{d\theta}{t_R}, \qquad (16)$$

и для трапецеидального входного сигнала при $t > t_0$

$$U_\sigma(t) = V\left\{1 - \frac{t_R}{t_0}\left[\exp\left(\frac{t_0}{t_R}\right) - 1\right]\exp\left(-\frac{t}{t_R}\right)\right\} \qquad (17)$$

Полученные формулы позволяют вычислить время распространения сигнала $t_\delta$, за которое отношение $U_\sigma/V$ достигнет заданной допустимой величины $\delta$. Примеры зависимостей $t_\delta$ от $t_0$ для трапецеидального входного сигнала приведены на Рис. 2. При $t_0 \ll t_\delta \ll t_{\sigma,R}$ из (14) и (17) следует, что $t_\delta \approx t_\sigma \delta^2/4$ и $t_\delta \approx t_R \delta$ соответственно.

Следует отметить, что обычно [2,4,5] затухание ступенчатого видеоимпульса характеризуется величиной $\Delta = 1 - U_f(t)/V$, где $U_f(t) = U(t, z_f)$ - скачок напряжения на фронте. Однако для вычисления $\Delta$ нужно иметь полное решение задачи Коши (1)-(3). Насколько нам известно, точное решение было получено только для линейной линии передачи без учета скин-эффекта [4,6]. Из него, как нетрудно показать следует, что

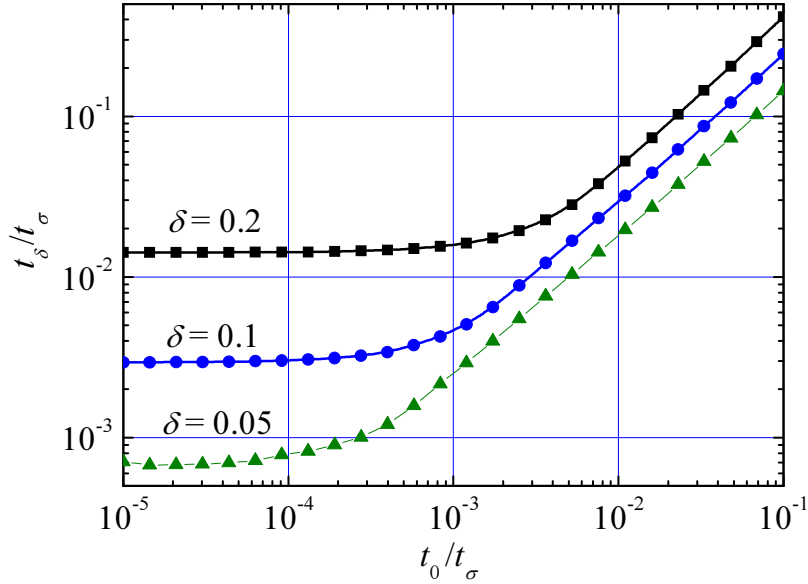

Рис. 2. Зависимости времен $t_\delta$ достижения отношением $U_\sigma/V$ величин $\delta$ от длительности фронта $t_0$ импульса на входе в линию, рассчитанные по формулам (9), (11) и (12).

$$\Delta(t) = 1 - \exp(-t/2t_R). \qquad (18)$$

тогда как

$$\delta(t) = 1 - \exp(-t/t_R). \qquad (19)$$

Эта последняя формула, разумеется, получается и из (17) в пределе $t_0 \to 0$ [2]. Таким образом напряжение на фронте $U_f$ уменьшается со временем в два раза медленнее, чем возрастает «омическое» падение напряжения $U_\sigma$. Причина этого очевидна: рост $U_\sigma$ приводит к уменьшение со временем тока $I(t,z)$ вдоль линии, так что второе слагаемое в (1) частично (в этом простейшем случае - ровно наполовину) компенсирует «омическое» поле $E_\sigma$ и замедляет уменьшение напряжения $U$ с ростом $z$. С учетом сильного скин-эффекта было получено приближенное решение [5,7], верное только при $t \ll t_\sigma$, когда

$$\Delta(t) \approx \frac{1}{2}\sqrt{\frac{t}{t_\sigma}}, \qquad (20)$$

то есть в этом случае $\Delta$ в 4 раза меньше $\delta$ (см. (14)).

На первый взгляд, предложенный метод оценки затухания видеосигнала менее нагляден, чем обычный. Однако у него есть два существенных преимущества. Во-первых, величина $\delta$, равная в первом приближении относительной потере мощности, сохраняет свой ясный смысл при любом соотношении между $t$ и характерным временем $t_0$ нарастания напряжения на входе в линию, а не только при $t_0 \to 0$. Во-вторых, при вычислении $\delta$ уравнение (2) не использовалось, поэтому полученные выше точные результаты справедливы и при сложной взаимосвязи между напряжением $U$ и плотностью заряда $Q$, когда необходимое для расчета $\Delta$ полное аналитическое решение задачи Коши (1)-(3) невозможно даже без учета скин-эффекта. В частности, это позволяет дать оценку величины затухания ударных электромагнитных волн в нелинейных линиях передачи [8].

---

[2] В этом пределе два первых слагаемых в правой части (7) отличны от нуля, но их сумма равна нулю, так как $dz_t/dt = 1/\sqrt{LC}$ и $U(t,z_t)/I(t,z_t) = \sqrt{L/C}$. Поэтому уравнение (8) и его решение (9) справедливо и для ступенчатого входного сигнала.